\documentclass[prl,amsmath,amssymb,dvips,twocolumn,showpacs]{revtex4} 
\usepackage{epsfig} 
\usepackage{bm} 
\usepackage{amsmath}
 
\newcommand{\be}{\begin{equation}} 
\newcommand{\ee}{\end{equation}} 
\newcommand{\bea}{\begin{eqnarray}} 
\newcommand{\eea}{\end{eqnarray}} 
 
\newcommand{\bp}{{\bf \psi}} 

\begin{document} 
 
\title{Effective 3-Body Interaction for Mean-Field and Density-Functional Theory}
 
\author{Alexandros Gezerlis$^1$ and G.~F. Bertsch$^{1,2}$ } 
\affiliation{$^1$Department of Physics, University of Washington, Seattle, WA 98195--1560 USA} 
\affiliation{$^2$Institute for Nuclear Theory, University of Washington, Seattle, WA 98195--1560 USA} 
 
\date {\today} 
 
\begin{abstract} 
Density functionals for nuclei usually include an effective 3-body 
interaction that depends on a fractional power of the density.
Using insights from the many-body 
theory of the low-density two-component Fermi gas, we 
consider a new, nonlocal, form for the energy functional that is
consistent with the Fock space representation of interaction operators.
In particular, there is
a unique spatially nonlocal generalization of the contact form of the interaction
that preserves the $\rho^{7/3}$ density dependence required by the
many-body theory.  
We calculate the ground state energies for particles in a harmonic 
trap using the nonlocal induced 3-body interaction,
and compare them to numerically accurate Green's Function Monte Carlo
calculations.  
Using no free parameters, we find that a nonlocality in the space domain
provides a better description of the weak-coupling regime
than the local-density approximation.
\end{abstract} 
 
 
\maketitle 

Self-consistent mean-field theory is the only practical tool to calculate
properties of heavy nuclei without region-specific parametrizations.
Two of the leading implementations, namely those following Skyrme
or Gogny, rely on interactions that depend on density, and
even on fractional powers of density, making them density-functional
theories rather than Hamiltonian theories.  We would like to
go beyond the density-functional theory as currently formulated,
and possibly back to effective Hamiltonian theories, 
for several reasons.  Correlation energies 
associated with 
angular momentum or particle number conservation are significant and 
need to be treated outside of the mean-field approximation.  Restoration
of good particle number can be carried out in many different ways in
a Hamiltonian theory, but density-functional theory can lead to inconsistencies
if the parameterization makes use of nonintegral powers of density
\cite{Duguet:2003,Robledo:2007,Duguet:2009}.  
Empirical evidence on nuclear compressional dynamics 
favors precisely such non-integral powers \cite{Blaizot:1980,Bender:2003},
although there have been attempts to keep only integral powers \cite{Baldo:2010,Erler:2010}.
Also, it is far  from clear that further accuracy can be achieved without 
dropping some of the
assumptions made, such as the Local Density Approximation (LDA).
These are our motivations for this Letter, to explore nonlocal alternatives
to the commonly used LDA energy functional.  This is not the first time
the question has been raised;  in a 2005 workshop\cite{INT05} the problem
was formulated:

``{\it Problem} 2. -- How can one replace in a nuclear density functional: (i) dependence on
momentum by dependence on density, or (ii) dependence on density by
dependence on momentum? The fact of life that nuclei are finite systems
composed of protons and neutrons must not be ignored, forgotten,
disregarded, neglected, or otherwise assumed irrelevant. The consequences of
the proposed replacements must be considered in the context of (i)
constructing functionals from first principles (e.g., how to replace the
Fermi momentum for the density), (ii) conserving symmetries (e.g., how to
construct an isospin-invariant density functional from microscopic results
for asymmetric matter), and (iii) restoring broken symmetries.''

Equation of state results for pure infinite neutron matter at densities $
\rho \geq 0.04$ fm$^{-3}$ have been commonly used to constrain Skyrme and
other density-functional approaches to large nuclei
\cite{Brown:2000,Stone:2003}.  It has recently become possible to use the
density dependence of the $^1\mbox{S}_0$ gap in low-density neutron matter
\cite{Gezerlis:2008} to constrain Skyrme-Hartree-Fock-Bogoliubov treatments
and especially their description of neutron-rich nuclei \cite{Chamel:2008}. 
At low densities it is possible to express the ground-state energy as an
analytically known function of $(k_F a)$, the product of the Fermi momentum
and the $s$-wave scattering length. Thus, finite systems of low density 
offer a good model for testing candidate effective 3-body interactions
and comparing with density-functional theory in the local density 
approximation.

As shown by Lee and Yang~\cite{Lee:1957}, the ground-state energy of a low-density Fermi gas with
short-range interactions can be expanded as a power series in the scattering
length $a$. The first 3 terms are: 
\be 
\label{LY} 
\frac{E}{N} = \frac{\hbar^2 k_F^2}{2 m}\left(
\frac{3}{5} + \frac{2}{3\pi}ak_F + \frac{4}{35\pi^2} \left ( 11 - 2 \ln2
\right ) \left ( ak_F \right )^2\right)~,  
\ee where $E/N$ is the energy per
particle and $k_F$ is the Fermi momentum.  

The corresponding energy density $\cal E$ expressed as a function of 
ordinary density $\rho$ ($=k_F^3/3\pi^2$) is
\be
\label{LYenerden} 
{\cal E} =  \frac{3\hbar^2}{10m} (3 \pi^2)^{2/3} \rho^{5/3} + \frac{\hbar^2 \pi a}{m}
\rho^2 + \frac{2 \hbar^2 a^2 3^{\frac{4}{3}} \pi^{\frac{2}{3}}}{35 m} 
(11 - 2 \ln 2) \rho^{\frac{7}{3}}.
\ee
The first and second terms
are just the kinetic and two-particle interaction energies of 
mean-field theory, using the scattering-length approximation to the
effective interaction.
The third term,
which we call an effective 3-body interaction, 
expresses an energy density functional that is proportional to $\rho^{7/3}$.

\begin{figure}[b]
\begin{center}
\includegraphics[width = 9cm]{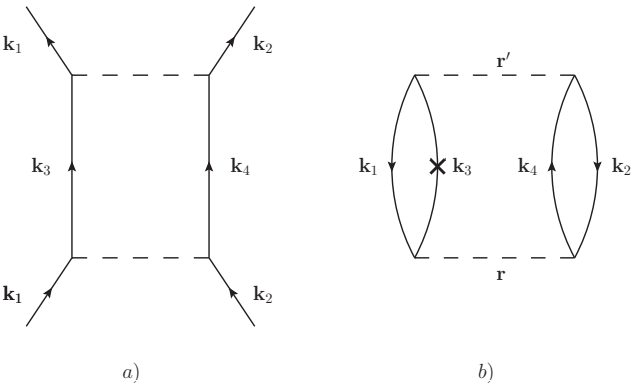}
\end{center}
\caption{\label{diagrams} a) A contribution to the effective two-particle
interaction in second-order perturbation.  b) The same perturbation
contribution, in a many-body context with orbitals $k_1, k_2$, $k_3$ occupied.}
\end{figure}

The origin of that term and its fractional density dependence may be
understood from the graphs in Fig.~\ref{diagrams}.  The low-density expansion makes use
of the scattering length, which is calculated by a ladder sum in the
two-particle channel.  One of the terms in that sum is shown in Fig. 
1a.  However, in the many-body context, shown in Fig. 1b, that
contribution should be excluded if either $k_3$ or $k_4$ is below the
Fermi momentum.  The third term in the Lee-Yang expansion is simply
subtracting out graphs of the form Fig. 1b, where the cross on the
particle line indicates that its momentum is below $k_F$ (and thus
it is not a proper Goldstone many-body graph).  In standard formulations
of many-body theory
\cite{Galitskii:1958, Bishop:1973, Fetter:2003, Hammer:2000} the graph
Fig. 1b is calculated as the integral
\be
\label{FGintegral} 
I_{LY} = V \int \frac{d^3 k_1}{(2 \pi)^3}  \frac{d^3 k_2}{(2 \pi)^3}   \frac{d^3
k_3}{(2 \pi)^3}  \frac{n_{k_1}n_{k_2}n_{k_3}}{k_1^2 +k_2^2 -
k_3^2-k_4^2} 
\ee
where $n_k = \theta(k_F -k)$ is the occupation number of the orbital $k$.

If we take the second-order perturbative expression for the
energy, but without assuming Fermi gas wave functions, and an
interaction of contact form the formula is:
\be
\label{eq:orbitals}
I = \sum_{1,2,3}^{occ} \sum_{4}
 \int d^3 r d^3r'  \frac{
\phi_{1{\mathbf r'}}^* \phi_{2{\mathbf r'}}^* \phi_{3{\mathbf r}}^* \phi_{4{\mathbf r}}^*
\phi_{1{\mathbf r}} \phi_{2{\mathbf r}} \phi_{3{\mathbf r'}} \phi_{4{\mathbf r'}}
}{E_1 + E_2 -E_3-E_4}
\ee
where $\phi_{i{\mathbf r}}$ is the orbital function of the $i$-th particle  
at position ${\mathbf r}$, and $occ$ signifies the fact that the states are occupied. 

The minimal generalization of a 3-body contact interaction is one that is
a function of two positions, ${\mathbf r}$ and ${\mathbf r'}$. Such an assumption leads
naturally to a precisely defined effective 3-body interaction that 
overcomes the problems associated with the $\rho^{7/3}$ density functional
but still reproduces the Lee-Yang functional dependence.  

The Fock-space representation of this operator is 
\be
\label{H3} 
\hat H_3 =
 f({\bf r, r'}) \sum_{\mbox{\boldmath$\sigma$}}\bp^\dagger_{\sigma_1 \bf
r'}\bp^\dagger_{\sigma_2\bf r'}\bp^\dagger_{\sigma_3\bf
r} \bp_{\sigma_4\bf r'}\bp_{\sigma_5\bf r}\bp_{\sigma_6\bf r}\,.
\ee
The two coordinates ${\bf r}$ and ${\bf r'}$ represent the two interaction
points in Fig. 1b.  The prescription to calculate expectation values with
this operator is to take the contractions in the Hartree-Fock ground state
that correspond to Fig. 1b, where the following triply contracted terms 
are the only ones to survive:
\begin{multline}
\label{C3}
\langle \hat H_3 \rangle =
\int d^3r d^3r' f({\bf r, r'}) \\
\sum_{\mbox{\boldmath$\sigma$}}\langle \bp^\dagger_{\sigma_1 \bf
r'}\bp_{\sigma_6\bf r}\rangle \langle  \bp^\dagger_{\sigma_2\bf r'} \bp_{\sigma_5\bf
r}\rangle \langle \bp^\dagger_{\sigma_3\bf
r} \bp_{\sigma_4\bf r'}\rangle \,
\end{multline}
Note that since the orbital $k_3$ is occupied, the graph is forbidden
(does not occur in a Goldstone expansion) and is marked with an x. 
Its contribution to the effective two-body interaction must 
be subtracted as part of the effective 3-particle interaction. 

The function $f({\bf r, r'})$ should be translationally and
rotationally invariant, i.e. it can only depend on $|{\mathbf r}-{\mathbf r'}|$.
Moreover, we want the functional that follows from 
the new operator to have the same dependence on
the density as the last term in Eq. (\ref{LYenerden}) 
(i.e. $\rho^{7/3}$) and we demand that it contain
$\hbar^2 a^2 / m$ so as to correspond to that term.
Finally, we see from Eq. (\ref{C3}) that 
$f({\bf r, r'})$ should have dimensions of $E L^3$.
The only way to satisfy all the above constraints is with an $f({\bf r, r'})$
of the form:
\be
\label{def3} 
f({\bf r, r'}) = \frac{\hbar^2 a^2}{m} \frac{C}{|{\mathbf r}-{\mathbf r'}|}~.
\ee

To obtain the dimensionless coefficient $C$, we demand that Eqs. \eqref{C3} \& \eqref{def3} reduce to the
Lee-Yang energy functional in the limit of uniform matter.  In that
limit, we can express the density matrices as
\be
\langle \bp^\dagger_{\sigma' \bf r'} \bp_{\sigma \bf r} \rangle =  
\int \frac{d^3 k}{(2\pi)^3} e^{i{\bf k} \cdot ({\bf r} - {\bf r'})} n_{k} \delta_{\sigma, \sigma'}
\ee
where $n_k$ is the Fermi gas occupation factor.
Carrying out the spatial integrals in Eq. \eqref{C3}, we have
\be
\label{H3integral} 
\langle \hat H_3 \rangle = 8 \pi L^3 C \frac{\hbar^2 a^2}{m} \int \frac{d^3 k_1}{(2 \pi)^3}  \frac{d^3 k_2}{(2 \pi)^3}   \frac{d^3
k_3}{(2 \pi)^3} \frac{n_{k_1}n_{k_2}n_{k_3}}{({\mathbf k}_3 - {\mathbf k}_1 - {\mathbf k}_2)^2} ~.
\ee
We match the result of this integral to the second-order Lee-Yang coefficient. This allows us to
go back to Eq. (\ref{H3}) and express the 3-body operator as a function of the two coordinates ${\mathbf r}$ and ${\mathbf r'}$,
with an operator structure that is guided by the terms in Eq. (\ref{C3}): 
\be
\label{H3result} 
\hat H_3({\bf r},{\bf r'}) = \frac{\hbar^2 a^2}{m} \frac{C}{|{\mathbf r}-{\mathbf r'}|} \sum_{\bf
\sigma}\bp^\dagger_{\uparrow \bf
r'}\bp^\dagger_{\downarrow\bf r'}\bp^\dagger_{\sigma\bf
r} \bp_{\sigma\bf r'}\bp_{\downarrow  \bf r}\bp_{\uparrow \bf r}
\ee
where
\be
C = \frac{64 \pi (11 - 2\ln2)}{3 (92 - 27\ln3)} \approx 10.336~.
\ee
Since we are dealing with a system of two fermionic species, in Eq. (\ref{H3result}) 
we have chosen to call them spin-up ($\uparrow$) and spin-down ($\downarrow$).
The two middle terms are free to take on either value of the spin.

A finite range 3-body effective interaction
like the one we propose in Eq.~(\ref{H3result}) is much softer than usually assumed.  
This fact could have important consequences on the density profile of different systems.
The price to be paid for
the conceptual clarity and microscopic derivation of our 3-body effective interaction
is the emergence of challenging computational issues. If $N$ is
the number of particles under study and $N_r$ is the number of amplitudes in the numerical vector 
respresenting an orbital, then for the usual way of doing Skyrme calculations, 
where the interaction depends only on the one-particle diagonal density, the latter is
of order $N N_r$. In contrast, the form of Eq. (\ref{H3result}) is of order $N^3 N_r^2$.

With a view to testing the effective interaction of Eq. (\ref{H3result}) we
consider the model problem of dilute fermions in a 
harmonic trap, as in Ref. \cite{Puglia:2003}.  We calculate the ground-state energies in density-functional
theory with the LDA functional, and with Eq. \eqref{H3result} replacing the Lee-Yang
term in the functional. For our purposes, we may take the orbitals
to be of the harmonic oscillator form $\phi({\bf r})=
P({\bf r})e^{-\nu r^2/2}$ where
$P({\bf r})$ is a polynomial and $\nu$ is a variational parameter.
We write down the variational energy of the system using the density as follows:
\be
E_{\nu} = T[\rho] + V_{\mathrm{ext}}[\rho] + 
\int d^3r  {\cal E}_I[\rho] + E_{II}.
\label{Enu}
\ee
In this expression, the kinetic energy is:
\be
T[\rho] = \sum_{i=1}^N \int d^3r \phi_i^{*}({\bf r}) \left ( -\frac{\hbar^2}{2m} \nabla^2 \right ) \phi_i({\bf r})
\ee
and the trap potential energy is:
\be
V_{\mathrm{ext}}[\rho] = \int d^3r \frac{1}{2} m \omega^2 r^2 \rho({\bf r})~.
\ee
The ${\cal E}_I$ is the second term in the Lee-Yang expansion Eq. (\ref{LYenerden}) and is simply:
\be
{\cal E}_I[\rho] = \frac{\hbar^2 \pi a}{m} \rho^2~.
\ee
For the last term, $E_{II}$, we choose two different forms: a) one following from the Lee-Yang expansion Eq. (\ref{LYenerden})
and b) one that corresponds to the new 3-body effective interaction given in Eq. (\ref{H3result}). These are:
\be
 E_{II}^{LY} = \int d^3 r {\cal E}_{II}^{LY} =
\frac{2 \hbar^2 a^2 3^{\frac{4}{3}} \pi^{\frac{2}{3}}}{35 m} (11 - 2 \ln 2) 
\int d^3 r \rho^{\frac{7}{3}}
\ee
and
\be
E_{II}^{H_3} = \frac{\hbar^2 a^2}{m} \sum_i^{\uparrow} \sum_j^{\uparrow \downarrow} \sum_k^{\downarrow} I_{ijk}
\ee
with
\be
I_{ijk} = 
\int d^3r d^3r'  \frac{C}{|{\mathbf r}-{\mathbf r'}|} \rho_i({\mathbf r},{\mathbf r'}) 
\rho_j({\mathbf r},{\mathbf r'}) \rho_k({\mathbf r},{\mathbf r'})~.
\ee
The equation for $E_{II}^{H_3}$ contains products of one-body density matrices. The sum
is taken in such a way that there are no repeated terms other than those allowed by the restriction
(clear by inspecting Eq. (\ref{H3result})) that only the $\uparrow \uparrow \downarrow$ and $\uparrow \downarrow \downarrow$ configurations are allowed.

We examine a system of 8 particles (4 spin-up and 4 spin-down). 
We choose 8 particles since this is the smallest non-trivial closed shell system:  
we find 128 terms in total (64 for $\uparrow \uparrow \downarrow$ 
and 64 for $\uparrow \downarrow \downarrow$). 
In Table \ref{table1} we show the results of variational minimizations of the functionals given
in Eq. (\ref{Enu}) in varying levels of sophistication. Shown are the energies of the system at the minima. 
We list the results 
when we keep a) only the kinetic energy and the external potential terms in Eq. (\ref{Enu}), 
b) the same two terms plus the next one (${\cal E}_I$), which is proportional to the scattering length $a$, 
 c) the above three terms plus the Lee-Yang highest order term (${\cal E}_{II}^{LY}$), proportional to $a^2$, 
and d) similarly to the previous case but with ${\cal E}_{II}^{LY}$ replaced by ${\cal E}_{II}^{H_3}$.

\begin{table}[b]
\caption{Results for the ground-state energy (in units of $\hbar \omega$) of a 
harmonically trapped system of 8 pairwise repulsively
interacting particles.
The interaction is at the dilute limit, $a \approx 0.2082 \sqrt{\hbar/m \omega}$.}
\begin{center}

\begin{tabular}{l r}
\hline
 & $E~[\hbar \omega]$\\
\hline
Kinetic + External Energy & 18.0 \\
DFT (LY LO) &  19.197 \\
DFT (LY LO + NLO) &  19.436 \\ 
DFT (LY LO + $H_3$ NLO) & 19.465 \\
GFMC & 19.485(1) \\
\hline
\end{tabular}

\end{center}
\label{table1}
\end{table}

To test the accuracy of the different approaches, we have performed
Green's Function Monte Carlo simulations, which have already been proved to be dependable in
describing the electron gas, light nuclei, and cold atoms \cite{Ceperley:1980,Pudliner:1997,Carlson:2003,Foulkes:2001,Gezerlis:2010}. 
As before, we study 8 trapped fermions, assuming a Hamiltonian 
of the form:
\begin{equation}
{\cal{H}} = \sum\limits_{k = 1}^{N}  ( - \frac{\hbar^2}{2m}\nabla_k^{2} + \frac{1}{2} m \omega^2 r_k^2)  + \sum\limits_{i<j'} v(r_{ij'})~.
\end{equation}
where $N$ is the total number of particles and $\omega$ is the trap frequency. The interaction between
the particles is taken to be of the modified P\"{o}schl-Teller type:
\begin{equation}
v(r) = v_0 \frac{\hbar^2}{m_r} \frac{\mu^2}{\cosh^2(\mu r)}~,
\label{eq:poeschl}
\end{equation}
where $\mu$ is a parameter that is related to the inverse of 
the effective range and
$v_0$ is a parameter we can use to tune the $s$-wave scattering length.
The scattering length for Table \ref{table1} is $a = 0.2082\sqrt{\hbar/m \omega}$.
For that case, the numerical calculations were performed taking
$v_0=10.28$ and $\mu^{-1} = 0.1$ in units of the oscillator
length $\sqrt{\hbar/m \omega}$.
Since the effective range is much smaller than the
oscillator length, this interaction is  appropriate to describe a
low-density system.  Also, we have taken the interaction to be repulsive
to avoid problems with superfluidity. 
We believe
that the errors associated with the fixed-node approximation are
inconsequential at the level of accuracy we are considering here.

\begin{figure}[t]
\begin{center}
\includegraphics[width = 9cm]{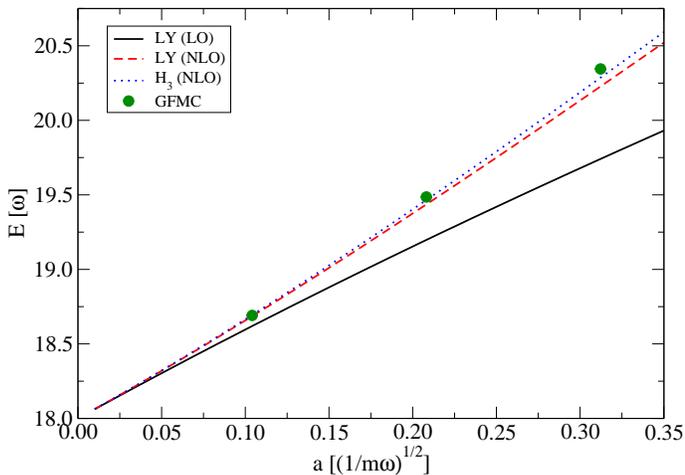}
\end{center}
\caption{\label{enerplot} The ground-state energy (in units of $\hbar \omega$) of a system of 8 particles 
plotted as a function of the scattering length (in units of $\sqrt{\hbar/m \omega}$).}
\end{figure}

We are now in a position to compare the various energies from Table \ref{table1} to the 
microscopic results. The Green's Function Monte Carlo result
is close to the density-functional theory calculation which includes all the terms in the Lee-Yang
expansion, but even closer to the results following from our new term $H_3$. 
We have also extended these calculations to different couplings, and gather 
our results in Fig.~\ref{enerplot}. At weak coupling, the new operator
leads to results that are identical with the Lee-Yang $a^2$ dependence, while as the coupling
gets stronger the $H_3$ contributions are more repulsive than in Lee-Yang (though they have the 
same power-law behavior), and thus provide
a more accurate description of the microscopic simulation. 
Importantly, our approach contains no free parameters: Eq. (\ref{H3result}) contains the parameter 
$C$, but that is matched to the appropriate Lee-Yang coefficient {\it for uniform matter}, 
while our subsequent calculation was performed for a finite particle number in a harmonic trap.

In summary, we have attempted to combine an awareness of the theory of the weakly interacting 
2-component Fermi gas with the desired behavior of density functionals, and have proposed a new form of the effective
3-body interaction that has a finite range. 
By including a new term of the form of Eq. (\ref{H3result}),
we are providing realistic density functionals with a way to match the analytically known 
behavior at very weak coupling.
We want to stress that to implement the proposed 3-body operator and functional 
what is required is only the one-body density matrix. This is computationally more challenging
than nuclear physics functionals based on the local density, but it is still much simpler
than approaches (such as coupled-cluster theory or the MBPT approximation of quantum
chemistry) that deal with correlations explicitly.

\begin{acknowledgments} 
We thank Ian Clo\"et and Aurel Bulgac for useful discussions.
Computations were performed at the National Energy
Research Scientific Computing Center (NERSC) and on the UW Athena cluster. 
This work was supported by DOE Grant Nos. DE-FG02-97ER41014 and DE-FG02-00ER41132.
\end{acknowledgments} 
 

 
\end{document}